# ULTRA-HIGH ANGULAR RESOLUTION BY GRAVITATIONAL MICROLENSING

MIKHAIL B. BOGDANOV

*Saratov State University, Astrakhanskaya st. 83,
410071 Saratov, Russia*

The problem of restoration of the source brightness distribution from an analysis of the stellar and AGNs microlensing light curves is investigated. In case of microlensing of stars by a point-mass lens as well as for caustic crossing events for binary lens the problem can be reduced to solution of the Fredholm integral equation of the 1st kind. Concrete form of the kernel of this equation depends on a type of the microlensing event. Assuming the circular symmetry of the stellar disk the search for radial brightness distribution can be carried out in the special compact sets of functions which correspond to the physics of the problem. These sets include the non-negative functions that are not increasing with increasing distance from the center of stellar disk and the upwards convex non-negative functions. The brightness distribution for the AGNs accretion disks is also circularly symmetric, but only in the locally co-moving frame. Therefore, the kernel of integral equation that determined the AGN microlensing light curve must take into account equally with the projection effect on picture plane the influence of relativistic effects. The search for solution of this equation can be carried out in the set of non-negative down convex functions. The results of analysis of microlensing light curves for the red giant MACHO Alert 95-30 and the A6 star MACHO 98-SMC-1 as well as the results of numerical simulations for the AGN microlensing observations are given.

## 1. INTRODUCTION

Many important problems of astronomy and astrophysics can be solved using the very high angular resolution observations in optical and infrared wave bands: the search for close binaries, the measurement of angular diameters of stars, the investigation of brightness distributions across the stellar disks, the study of fine spatial structure of active galactic nuclei (AGN). The high angular resolution is achieved in different ways to date. The speckle interferometry techniques, modern versions of the Michelson interferometer, and lunar occultation observations are applied to solution of this problem. The resulting resolution for modern ground based observations is of order 0."001.

A new tool to achieving of high angular resolution has been realized recently - the observations of gravitational microlensing. In role of target bodies in this case can be stars of our Galaxy or some another nearest stellar systems as well as the accretion disks of the distant AGNs. In role of gravitational lens are possible the usual stars or dark massive compact objects of halo of Galaxy (MACHO). The nature of these objects is still unknown. The gravitational lens can be considered frequently as the point-mass lens which possess of a spherically symmetric potential. The flux from a lensed source is amplified in this case depending on its angular distance from the lens. In rare observed events there are the binary gravitational lens. In this case the character of the gravitational potential creates caustics in the source plane. The crossing of a caustic leads to sharp variation of flux from the lensed source. High magnification events in multiple images of a lensed quasar due to gravitational microlensing by stars in a lens galaxy are also connected with the complex caustic system which is created by total gravitational potential (Schneider *et al.*,1992; Zakharov, 1997). The possible angular resolution of microlensing observations is measured by the value of order or less microsecond of arc.

The aim of our paper is investigation of the problem of restoration of the source brightness distribution from an analysis of the microlensing light curves.

## 2. MICROLENSIG BY POINT-MASS LENS

The amplification of a point-mass gravitational lens is determined by the value of Einstein cone angular radius $p_0$. This value is given by

$$p_0^2 = 4r_g D_{SL} / D_{OL} (D_{SL} + D_{OL}), \qquad (1)$$

where $r_g = GM/c^2$ is the gravitational radius of lens with mass M, G is the gravitational constant, c is the velocity of light, $D_{SL}$ is the distance from a lensed source to lens, and $D_{OL}$ is the distance from the lens to observer.

Let ds is an elementary area on a non-coherent radiate source. In absence of a gravitational lens this area is correspond to the elementary solid angle $d\omega$ and radiates the flux $dI_0$. At presence of lens the observed flux from this area dI can be written as $dI = A(p_s)dI_0$. The coefficient of amplification is equal

$$A(p_s) = \left( \sqrt{1 + 4p_0^2 p_s^{-2}} + 1/\sqrt{1 + 4p_0^2 p_s^{-2}} \right)/2, \qquad (2)$$

where $p_0$ is determined by equation (1), and $p_s$ is the angle between the lens and center of the area ds (Schneider *et al.*,1992; Zakharov, 1997).

If the source is circularly symmetric in picture plane with the radial brightness distribution b(r) in the polar coordinate system $(r,\varphi)$, then observed flux I(p) is depend only on an angle p between direction on center of the source and the lens. Integration over of all elementary areas gives the sample of microlensing light curve at the present moment

$$I(p) = \int_0^{2\pi} d\varphi \int_0^{\infty} A(\sqrt{p^2 + r^2 - 2pr\cos\varphi}\,)b(r)rdr \quad . \tag{3}$$

The equation (3) can be written in more compact form using the kernel

$$K(p,r) = r\int_0^{2\pi} A(\sqrt{p^2 + r^2 - 2pr\cos\varphi}\,)d\varphi \quad . \tag{4}$$

In these designations the connection of the observed microlensing light curve I(p) with the radial brightness distribution b(r) is given by

$$I(p) = \int_0^{\infty} K(p,r)b(r)dr \quad . \tag{5}$$

Thus, the search for b(r) is the inverse problem for the integral equation (5).

In case of microlensing observations the light curve is registered as a function of time t . If the relative motion of lens and source is straightforward and uniform, then

$$p^2(t) = p_m^2 + V^2(t - t_m)^2 \quad , \tag{6}$$

where $p_m$ is the minimal value of angular distance at moment $t_m$ , and V is the angular velocity. The values $t_m$ and V are the free parameters of our problem.

The equation (5) is a Fredholm integral equation of the 1st kind and the inverse problem of restoration of b(r) belongs to class of the ill-posed problems which require to utilize *a priori* information about solution (Goncharsky A.V. et al., 1985). Bogdanov and Cherepashchuk (1996) showed that the brightness distribution b(r) can be restored from the microlensing light curve I(p) as a function of an angle, which is measured in units of the lens's Einstein cone angular radius. Later this problem has been examined also by Hendry et al. (1998). Unfortunately, the sensibility of solution of this inverse problem to the influence of a brightness distribution is low usually. A rare situation when the lens is projected on the stellar disk is especially interesting. Such case has been observed by MACHO and GMAN collaborations in the event MACHO Alert 95-30 for a red giant of the Galactic bulge (Alcock et al., 1997). Figure 1 shows a part of the microlensing light curve for this event. We carried out the search for b(r) in the class of non-negative functions that are not increasing with increasing distance from the center of stellar disk (Bogdanov and Cherepashchuk, 1999). The result is shown in the figure 2. The restored profile b(r) agrees qualitatively with results of the extended atmosphere model calculations for the red giants.

## 3. MICROLENSING BY CAUSTIC OF BINARY LENS

A difficulty of an analysis of the observational data of microlensing for a binary gravitational lens by a model fitting method is connected with a big number of free model parameters (Albrow et al., 1999). It can be shown that for the microlensing by a fold caustic the shape of b(r) can be determined without a knowledge of majority lens parameters. Further we shall examined only this case of microlensing as more probable for the real observations.

The angular size of observed stars of the bulge of Galaxy or the Magellanic Clouds is very small. Therefore, we may neglect of a caustic curvature and regard the fold caustic as a straightforward line. In this case the picture of lensing will be depend only on one-dimensional strip brightness distribution B(x) in the direction of x axis that is perpendicular to caustic. B(x) is connected with two-dimensional brightness distribution b(x,y) by integral equation

$$B(x) = \int_{-\infty}^{\infty} \int_{-\infty}^{\infty} b(\mathbf{x}, y)\, \delta(\mathbf{x} - x)\, d\mathbf{x}\, dy \quad , \tag{7}$$

where $\delta(x)$ is the Dirac function. For a circularly symmetric source b(r) is connected with B(x) by the Abel integral equation

$$B(x) = \int_{x}^{\infty} 2b(r)\, r\, dr / \sqrt{r^2 - x^2} \quad . \tag{8}$$

Let we observe the second caustic crossing which leads to sharp drop of the flux. If $\xi$ is the angular distance of point of B($\xi$) from its center and x is the angular distance of source center from caustic then the coefficient of amplification can be written as (Schneider *et al.*,1992; Zakharov, 1997)

$$A(x,\xi) = A_0 + k\, H(\xi - x) / \sqrt{x - x} \quad , \tag{9}$$

where H(x-$\xi$) is the Heaviside step function (H = 0 for negative and H = 1 for non-negative values of its argument), $A_0$ and k are certain constants for the given microlensing event. The observed microlensing light curve as a function of distance x is given by the convolution integral equation

$$I(x) = A(x) * B(x) = \int_{-\infty}^{\infty} A(x - \mathbf{x})\, B(\mathbf{x})\, d\mathbf{x} \quad . \tag{10}$$

The precision of *a priori* estimation of the constant $A_0$ and k is poor enough. The product $A_0 I_0$, where $I_0$ is the full flux from the source, can be evaluated from an analysis of a section of the microlensing light curve which placed right away of drop of I(x). If we subtract value of this constant level from observational data, then B(x) can be restored from equation (10) with an accuracy up to any constant factor. The distance $x = V(t - t_0)$, where V is the relative velocity of the caustic, t is the time, and $t_0$ is the time of the source center crossing. When a value of V is unknown the linear scale of B(x) can be determined also with an accuracy up to any constant factor. Therefore, if we adopt in equation (9) $A_0 = 0$, $k = 1$, and admit $V = 1$ then only the shape of the brightness distribution can be found from the caustic microlensing observations. In case of stars, for example, this limited information is very important for the comparison with data of atmosphere model calculations.

For circularly symmetric source we may obtain, combining of equations (8), (9), and (10), the integral equation that connects the observed light curve I(x) with the radial brightness distribution b(r)

$$I(x) = \int_0^\infty S(x,r) b(r) dr \quad . \tag{11}$$

The kernel of this equation S(x,r) defines the process of amplification of flux from the source by fold caustic. The form of equation (11) is completely analogous to form of equation (5). The inverse problem for this equation has only one free parameter - the correction of null-punct of x axis $\Delta x$. It is important that $\Delta x$ can be found independently from solution of equation (10) as a shift of the center of contour B(x).

As an example, we present the results of analysis of the caustic microlensing curve for A6 spectral type star in Small Magellanic Cloud observed by PLANET collaboration in event 98-SMC-1 (Albrow et al., 1999). Figure 3 shows by circles the samples of the observed curve disposed not far from its maximum. The search for b(r) has been carried out in the class of the upwards convex nonnegative functions. Figure 4 shows the restored shape of b(r) together with the curve of non-linear law of limb darkening for the A6 star.

## 4. MICROLENSING OF ACCRETION DISK OF AGN

The observations of high magnification events in multiple images of a lensed quasar due to gravitational microlensing by stars in a lens galaxy allow to obtain an information about a quasar central engine - the accretion disk, which surrounds a super massive black hole in galactic nucleus (Grieger et al., 1991; Mineshige & Yonehara, 1999; Agol & Krolik, 1999). The problem also reduces to solution of integral equation of type of the equation (11), however, b(r) regards as the brightness distribution in the locally co-moving frame of accretion disk. In contrast to brightness distribution observed by external distant observer, the profile of b(r) is circularly symmetric. The kernel of integral equation must take into account in this case equally with the amplification by a caustic also the inclination of the disk to line of sight and the influence of relativistic effects.

In general case, the brightness distribution across the disk in locally co-moving frame can be restored only as a two dimensional function of frequency and radius (Agol and Krolik, 1999). Consequently, the accurate analysis can be carried out only for multi-color photometry or spectral observational data. It is necessary the knowledge of a disk radiation model for an analysis of a single-frequency microlensing light curve. We chose the standard model of thin relativistic accretion disk with a constant accretion rate and no advection of heat derived by Shakura (1972) and Shakura and Sunyaev (1973). For this model the energy generation per unit area Q(r) as a function of radius r is given by

$$Q(r) = \frac{3GM\dot{M}}{4\pi r^3} R_R(\eta) \quad , \qquad (12)$$

where M is the mass of the black hole, $\dot{M}$ is the accretion rate, $\eta = r/r_g$, $r_g = GM/c^2$, and $R_R(\eta)$ is a correction factor that take into account the influence of relativistic effects (Novikov and Thorne, 1973; Page and Thorne, 1974; Krolik, 1999).

For calculation of the brightness distribution b(r) we make several simplifying assumptions: (I) the Kerr's black hole has the maximal spin a/M = 0.998 (Thorne, 1974); (II) the accretion disk is geometrically thin and optically thick; (III) the disk is flat and lies in the equatorial plane of the black hole; (IV) the gas follows prograde circular orbits outside the marginally stable radius; (V) the gas emits isotropically in its rest frame; (VI) the local thermodynamic equilibrium is valid for each elementary area of the disk. The mass of black hole adopts to be equal $10^8$ $M_\odot$ and the accretion rate $\dot{M}$ = 1 $M_\odot$/year. The total flux from face-on disk in the locally co-moving frame is adopted to be equal to unit. The angle of inclination of disk axis to line of sight i is 30º and the angle between the major axis of projection of the disk on picture plane and the caustic motion direction $\alpha$ is 20º.

Our calculations showed that the contribution of central region of disk in total flux is low for observations in visible and infrared. This contribution decreases additionally for the distant external observer by the gravitational red shift. Therefore, the influence of region with r ≤ $r_0$ = 10 $r_g$ can be neglected and we can formally adopt inside this region b(r) = 0.

The brightness of an elementary area of the accretion disk is depend on the relativistic effects. The frequency shift of a photon emitted by atom of gas which follows a prograde circular orbit in the equatorial plane of the Kerr's black hole g = $\nu_0/\nu_e$, where $\nu_0$ is the observed frequency and $\nu_e$ is the emitted frequency, can be determined by means of the precise formulae (Chandrasekhar, 1983; Novikov and Frolov, 1986). It is known that the bending of photon trajectories is very strong in central region of an accretion disk where arises a short wavelength radiation. Therefore, it is necessary the precise calculation of general relativity effects in Kerr metric for a correct interpretation of X-ray and UV observations of the AGN. For observations in visible or infrared the problem can be appreciable simplified.

The calculations of photon trajectories in Kerr metric show that the effects of curvature of the frame are essential only in the central region of the accretion disk for $r \leq r_0$ ( Cunningham, 1975; Laor, 1991; Karas et al., 1992; Rauch and Blandford, 1994; Bromley et al., 1997; Pariev and Bromley, 1998; Zakharov and Repin, 1999). Therefore, we can consider approximately that a photon trajectory is the straightforward line outside the central region of disk and the solid angle of an elementary area on the disk surface coincident with its value for the flat frame. The brightness observed by an external distant observer in any point of an accretion disk $b_0(\boldsymbol{n}_0)$ is connected with the brightness in the locally co-moving frame $b(\boldsymbol{n}_e)$ by famous relation: $b_0(\boldsymbol{n}_0) = g^3 b(\boldsymbol{n}_e)$.

It can be shown in our assumptions that the observed microlensing light curve $I_\nu(x)$ is connected with the brightness distribution $b_\nu(r)$ for more probable case of a fold caustic lensing by integral equation

$$I_{\boldsymbol{n}_0}(x) = \int_{r_0}^{\infty} P(x, r, g, i, \boldsymbol{a}, \Delta x) b_{\boldsymbol{n}}(r) \, dr \quad . \tag{13}$$

The kernel $P(x,r,g,i,\alpha,\Delta x)$ is depend equally with the coordinates and frequency shift g also on the angle of inclination of disk axis to line of sight i, the angle between the major axis of projection of the disk on picture plane and the caustic motion direction $\alpha$, and the correction of null-punct of x axis $\Delta x$. The frequency shift caused by the relativistic effects transfers to the observational frequency $\nu_0$ in different areas of the accretion disk the different samples of the initial radiation spectrum. Therefore, it is necessary the knowledge of a disk radiation model (in an emergency the relative energy distribution) for an analysis of a single-frequency microlensing light curve. This factor must be taken into account in calculation of the kernel $P(x,r,g,i,\alpha,\Delta x)$. As a consequence of it, solving the inverse problem for equation (13) we shall understand under $b_\nu(r)$ the brightness of the disk in locally co-moving frame at the radius r for any frequency $\nu$ which is connected with the observational frequency $\nu_0$ by average value of the factor g weighted with the caustic amplification coefficient (9). In general case the cosmological red shift also must be considered, but for the present we ignore its influence.

The search for solution of the ill-posed inverse problem for integral equation (13) can be carried out in the class of non-negative down convex functions (Goncharsky at al.,1985). The quantities i, $\alpha$, and $\Delta x$ are free parameters of the problem. As it be discussed above, if the caustic parameters from equation (9) are unknown then only the shape of the brightness distribution can be found from microlensing observations.

For estimation of the ability of this method we carried out the numerical simulations. We used the brightness distribution in V photometric band for our model of the accretion disk as an initial function. The samples of the microlensing curve have been distorted by addition of the Gauss stochastic noise with the null mean and the standard deviation $\sigma = 0.01$ of maximal flux. These sam-ples are sown in figure 5 depending on the distance from the caustic x, measured in units of gravitational radius $r_g = GM/c^2$.

The search for the brightness distribution has been carried out in the class of non-negative down convex functions using the updated computer code of Goncharsky at al. (1985) which minimizes $\chi_N^2$ by the projection conjugate gradient method. Figure 6 shows the samples of the restored distribution b(r). The radial distance r is also measured in units of $r_g$. The initial profile of brightness distribution is given by the solid line. It is clear that this distribution is reproduced quite well.

## 5. CONCLUSION

The results of our investigations show that observations of the gravitational microlensing is very powerful method to study of the angular structure of the lensed sources. The problem of the data analysis can be reduced to solution of the Fredholm integral equation of the 1st kind. Concrete form of the kernel of this equation depends on a type of the microlensing event. The extension of the volume of *a priori* information about solution of the ill-posed inverse problem by means of search for the brightness distributions in the special compact sets of functions which correspond to the physics of the events allows to enhance the precision and stability of the solution and to achieve the ultra-high angular resolution. The analysis of the microlensing light curves of stars can be carried out easy and unambiguous as for the point-mass lens equally as for the binary gravitational lens. The restored brightness distributions for stars in microlensing events observed by MACHO, GMAN and PLANET collaborations are agree qualitatively with results of modern stellar atmosphere model calculations.

The derivation of any information about the accretion disk of the AGN from the microlensing observation is a difficult problem. In general case, the brightness distribution across the disk in locally co-moving frame can be restored only as a two dimensional function of frequency and radius. Consequently, the accurate analysis can be carried out only for multi-color photometry or spectral observational data. For the observations in visible or infrared the influence of the relativistic effects is comparatively low and can be taken into account with any simplifying approximations. In this case, however, it is necessary the knowledge of the relative energy distribution of the disk radiation for an analysis of a single-frequency microlensing light curve. The circularly symmetric radial brightness distribution for the accretion disk in locally co-moving frame can be restored by search for solution of the main integral equation in the class of non-negative down convex functions.

This work was partially supported by grants of Russian State Scientific Program "Astronomy" and Program "Universities of Russia".

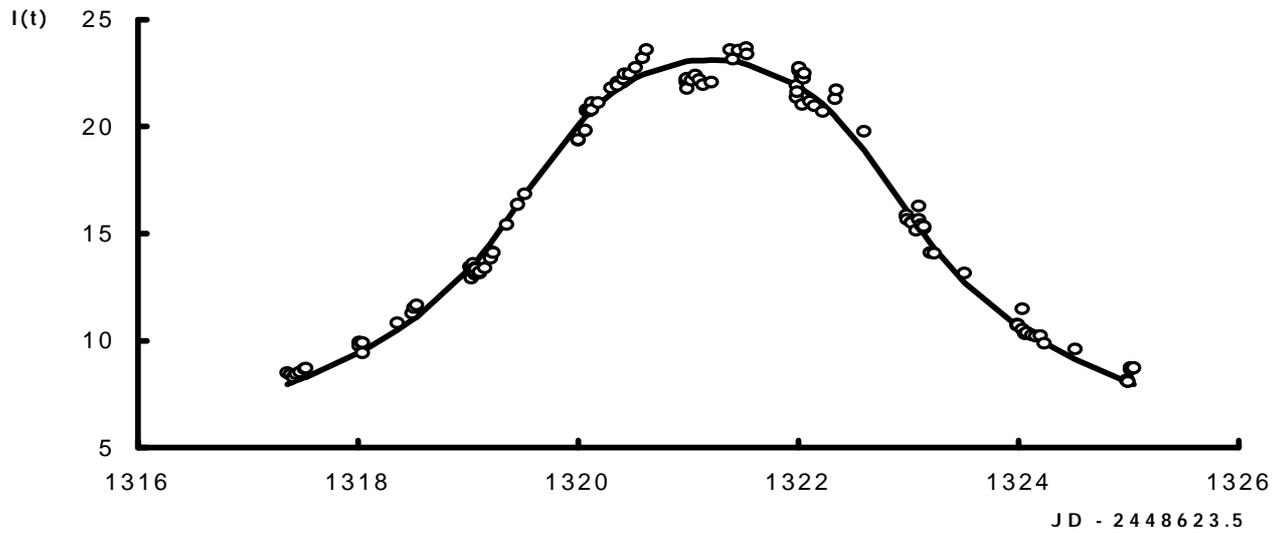

Fig.1. The samples of microlensing light curve for red giant MACHO Alert 95-30 observed by MACHO and GMAN collaborations (circles). The solid line is the microlensing curve for restored brightness distribution.

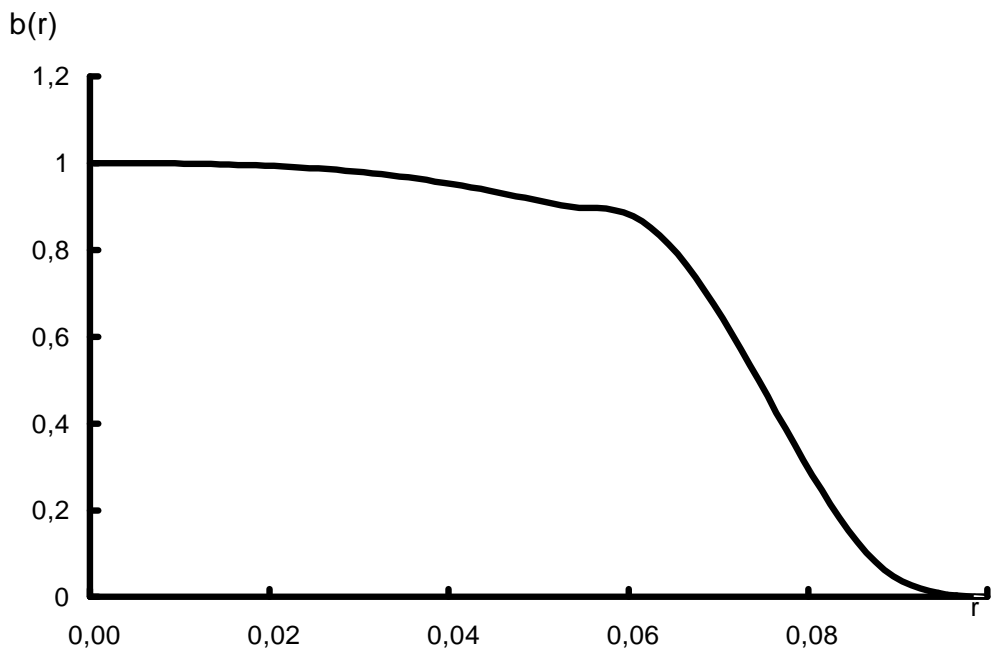

Fig.2. The radial brightness distribution across the disk of red giant MACHO Alert 95-30 restored in the class of non-negative functions that do not increase with increasing distance from the center of stellar disk as the function of angle measured in units of the lens's Einstein cone angular radius. The estimated angular radius of this star is 0."0000315 .

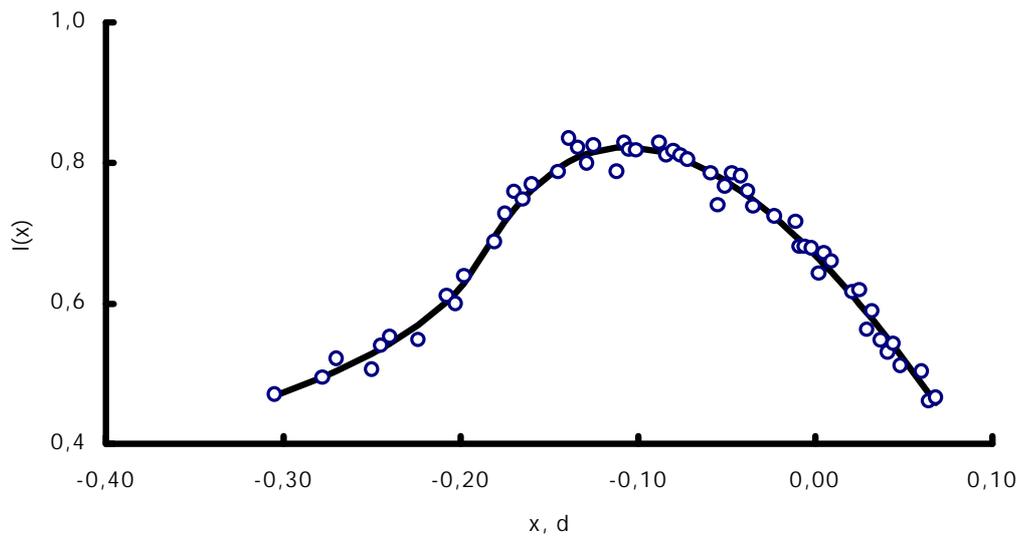

Fig.3. The samples of the light curve for microlensing by caustic of the binary lens in event MACHO 98-SMC-1 observed by PLANET collaboration as the function of time (in days) from the moment of center disk crossing (circles). The solid line is the microlensing curve for the restored brightness distribution.

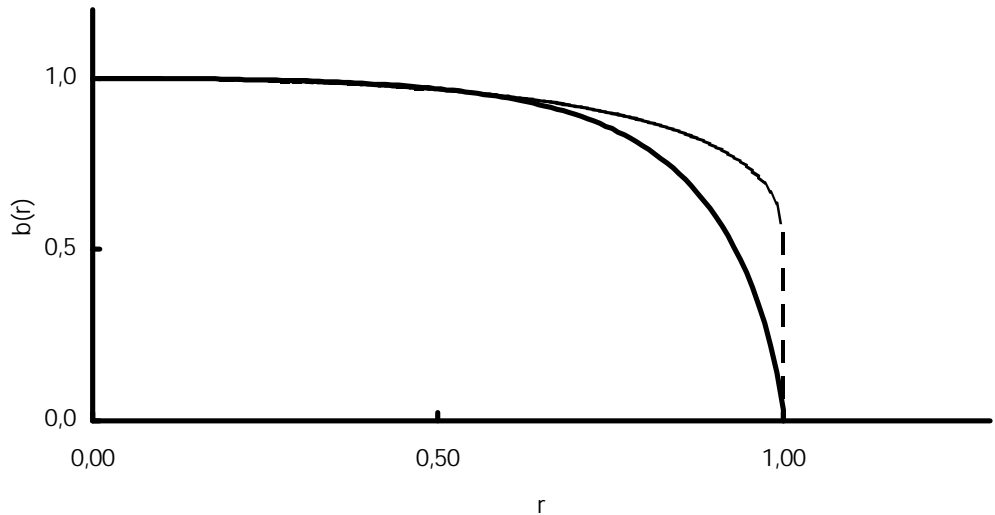

Fig.4. The shape of the radial brightness distributions across the disk of star MACHO 98-SMC-1 restored in the class of non-negative upwards convex functions (solid line) and the nonlinear law of limb darkening for atmosphere model of the star of spectral type A6 (dashed line). The estimated angular radius of this star is 0.″000000089 .

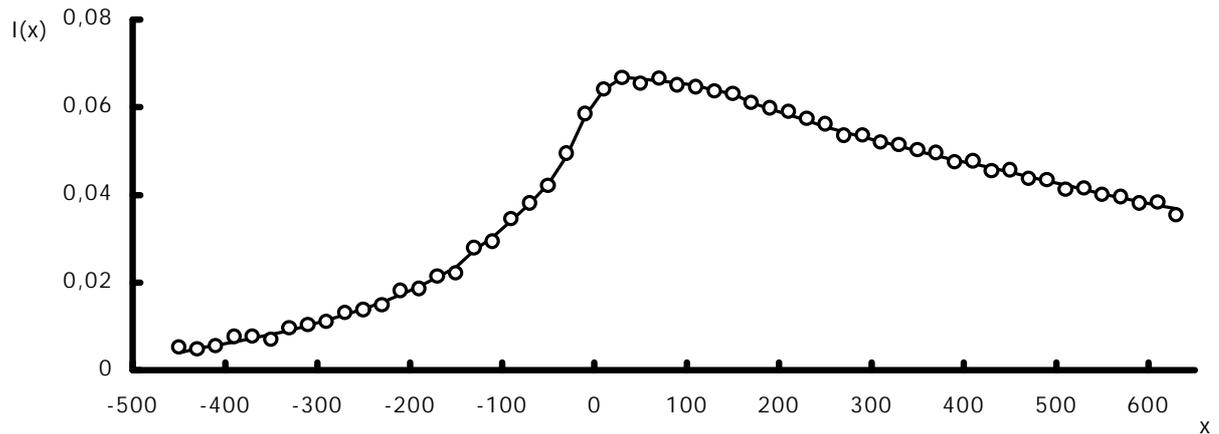

Fig.5. The samples of the simulated microlensing light curve of accretion disk of the AGN in V band with an additive noise level 1 % of the maximal value (circles) as a function of distance from the caustic (in units of the gravitational radius $r_g = GM/c^2$ ). The curve which correspond to the restored radial brightness distribution is showed by solid line.

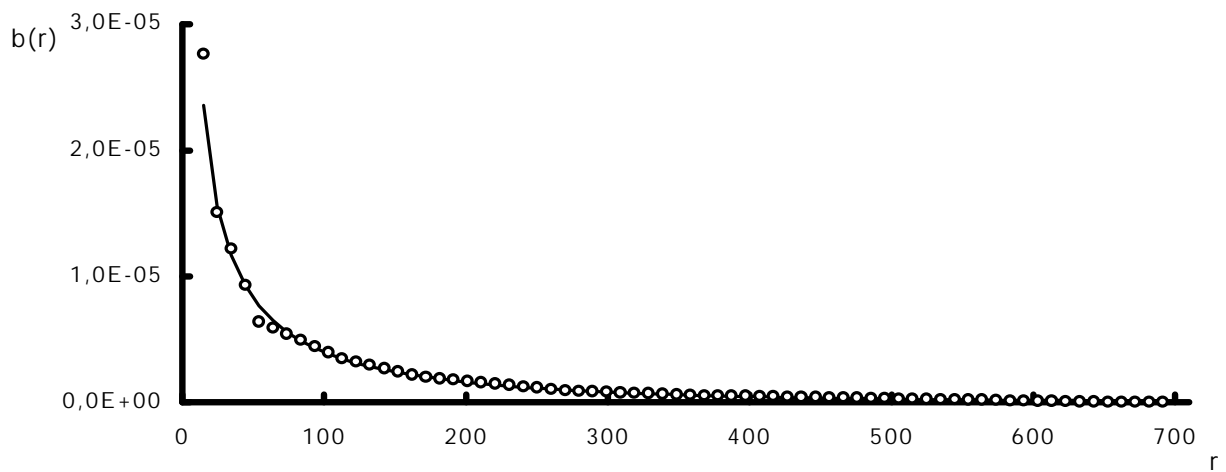

Fig.6. The radial brightness distribution for the accretion disk in the locally co-moving frame restored from the samples of the microlensing light curve of fig. 5 in the class of non-negative down convex functions (circles). The intrinsic profile is showed by solid line. The typical angular size of a quasar optical emission region is of order 0."00000001 .